\def\L{{\cal L}}
\def\Lb{{\Lambda}}
\def\tp{{t^{\prime}}}
\def\bp{{b^{\prime}}}
\def\taup{{\tau^{\prime}}}
\def\nutaup{{\nu_{\tau^{\prime}}}}
\def\vev{{\frac{{\it v}_u}{{\it v}_d}}}
\def\vevskew{{{\it v}_u/{\it v}_d}}
\documentstyle[12pt]{article}
\begin{document}
\begin{flushright}
(May 1994)
\end{flushright}
\centerline{\large \bf Quark-lepton flavor democracy and }
\centerline{\large \bf the non-existence of  the fourth  generation}
\vspace{1.2cm}
\begin{center}
{{\bf G.~Cveti\v c} \\
Inst.~f\"ur Physik, Universit\"at Dortmund, 44221 Dortmund,
Fed.~Rep.~of Germany \\[0.8cm]
{\bf C.~S.~Kim} \\
Department~of Physics, Yonsei University, Seoul 120-749, Republic of Korea}
\end{center}
\vspace{1.2cm}
\centerline{\bf Abstract}

In the Standard Model with two Higgs doublets (type II), which has
a consistent trend to a flavor gauge theory and its related flavor democracy
in the quark and the leptonic
sectors (unlike the minimal  Standard Model) when the energy of the probes
increases,
we impose the mixed quark-lepton flavor democracy at high ``transition'' energy
and
assume the usual see-saw mechanism, and
consequently find out that the existence of the fourth generation of
fermions in this framework is practically ruled out.\\
PACS number(s): 12.15.Ff, 12.15.Cc, 11.30.Hv

\newpage

The number of the light neutrino species with masses below $M_Z/2$ has been
measured very accurately at LEP~\cite{LEP}, and it is only three.
However, we believe this measurement does not completely exclude the
existence of the fourth generation, if the neutrino of this
generation is, for as yet unknown reasons, very heavy (presumably
heavier than $M_Z/2$). In this paper we would like to examine the
question of the non-exsistence of the fourth generation in the
framework of flavor gauge theory and its related flavor democracy
(FD)~\cite{gccsk}.

The notion of  flavor democracy  for quarks (q-q FD) at low (``physical'')
energies is well-known by now~\cite{fdle}. It means that a fermion
flavor basis\footnote{
This is a basis in which $\Psi_L$ ($\Psi^T =
(\psi_u, \psi_d)$) transforms as an iso-doublet under the $SU(2)_L$.}
exists in which the Yukawa couplings to $u_R$ quarks (and separately
to $d_R$ quarks) are all equal and that there is no Cabibbo-Kobayashi-Maskawa
(CKM)
mixing.  The FD in the leptonic sector (charged leptons and Dirac neutrinos)
is defined analogously. In reality, fermions at low energies manifest
the flavor democratic structure only to a first approximation. However, at
increasing
energy of probes, the minimal Standard Model (MSM) and the closely related
``type I'' Standard Model with two Higgs doublets (2HDSM(I)) have a clear trend
away from FD, while the usual ``type II'' Standard Model with two Higgs
doublets
(2HDSM(II)) has a consistent trend to FD in all quark and leptonic
sectors (q-q FD and $\ell$-$\ell$ FD)~\cite{gccsk}.

In the mass basis, neglecting the light first generation,
the trend to FD in the quark sectors (q-q FD) means:
\begin{equation}
 \frac{m_s}{m_b} \ , \ \frac{m_c}{m_t} \ , \ (V_{ckm})_{cb}
\longrightarrow 0 \qquad \mbox{as} \ E \rightarrow \Lb_{pole} \ ,
\label{FDqq}
\end{equation}
and analogously in the leptonic sectors ($\ell$-$\ell$ FD). Here,
$\Lb_{pole}$ is the Landau pole where the Yukawa couplings blow up.

The motivation behind these notions is the following. One could assume
that the Standard Model at some high ``transition'' energy $E_{trans.} \sim
\Lb_{pole}$ is replaced by a new strongly interacting physics with
almost flavor-blind forces, where the tiny deviation from the
``flavor-blindness'' is provided by some as yet unknown mechanism
(e.g.~a mechanism having its origin in string theory, etc.) and/or possible
radiative corrections near  $E_{trans.}$. In such a
framework, the 2HDSM(II) would be definitely favored as the low energy
theory, in comparison to the MSM or to its closely related
2HDSM(I)~\cite{gccsk}.
Furthermore, the Higgs sector could possibly be explained through the
mechanism of condensation of the neutral Higgses~\cite{shk} near
$E_{trans.}$, allowing realistically low values for $m_t^{phy}$
($m_t^{phy} < 200$ GeV for $E_{trans.} \sim \Lb_{pole} \ll \Lb_{Planck}$,
for a large range of values of the ratio of the vacuum expectation values:
$\vevskew \stackrel{<}{\sim} 1.75$~\cite{gccsk},\cite{moriond}), {\it unlike}
the condensation mechanism leading to the MSM~\cite{bhl}.
The entire results concerning the trend to FD can, however, be regarded
independently of the motivation outlined above.

If we want to have the number of degrees of freedom at $E_{trans.}$
($\sim \Lb_{pole}$) in the Yukawa sector of the 2HDSM(II) additionally
reduced, we can {\it impose} the mixed quark-lepton flavor democracy (q-$\ell$
FD)
\begin{equation}
  m_t \simeq m^D_{\nu_{\tau}} \ , \ m_b \simeq m_{\tau}
\qquad \mbox{at} \ E \simeq \Lb_{pole} \ ,
\label{FDql1}
\end{equation}
which would leave us at high ``transition'' energy with basically only
two Yukawa couplings ($g_{up}$ and $g_{down}$; $|g_{up}| \gg |g_{down}|$),
and would provide\footnote{
Since $m_b^{phy} \ll m_t^{phy}$, the condition $m_b \simeq m_{\tau}$
at $E \simeq \Lb_{pole}$ is automatically satisfied (``pull-up'' effect)
through the condition $m_t \simeq m^D_{\nu_{\tau}}$ at $E \simeq \Lb_{pole}$.}
us with the Dirac neutrino mass $m^D_{\nu_{\tau}}$ (at $E=1$ GeV) and
$\Lb_{pole}$,
for a chosen $m_t^{phy}$ and $\tan \beta \equiv \vevskew $.
The see-saw mechanism would subsequently furnish us with an estimate
(upper bound  ({\it u.b.})) for the resulting physical neutrino
mass~\cite{gccsk}
\begin{equation}
M_{_{Majorana}} \stackrel{>}{\sim} E_{trans.} \simeq \Lb_{pole} \quad
\Longrightarrow \quad
m^{phy}_{\nu_{\tau}} \simeq \frac{(m^{D}_{\nu_{\tau}})^{2}}
{M_{Majorana}} \stackrel{<}{\sim} \frac{(m^{D}_{\nu_{\tau}})^{2}}{\Lb_{pole}}
=(m^{phy}_{\nu_{\tau}})^{u.b.} \ .
\label{see-saw1}
\end{equation}

Here we would like to raise the question whether in this flavor
democracy--favored
2HDSM(II) model the imposed q-$\ell$ FD would be compatible with the existence
of the
fourth generation of heavy fermions ($\tp , \bp$), ($\nu^D_{\taup} , \taup$):
\begin{equation}
  m_{\tp} \simeq m^D_{\nutaup} \ , \ m_{\bp} \simeq m_{\taup}
\qquad \mbox{at} \ E \simeq \Lb_{pole} \ .
\label{FDql2}
\end{equation}
Through the application of the
1-loop renormalization group equations (RGEs) for the Yukawa couplings
in the $\overline{MS}$ scheme (see Appendix),
this condition  (4), together with the known masses of the third generation
fermions (we neglect the influence of the light fermions of the first
and the second generations - a very good approximation),
yield  a one-to-one  correspondence
\begin{equation}
m_t^{phy} \ , \ \tan \beta (\equiv \vev ) \ , \ m_{\tp}^{phy} \ , \
m_{\bp}^{phy}
\qquad \longleftrightarrow \qquad
\Lb_{pole} \ , \ m_{\taup}^{phy} \ , \ m^D_{\nutaup} \ ,
\label{corresp}
\end{equation}
where $m^D_{\nutaup}$ is the Dirac mass of the heavy neutrino at
$E = 1$ GeV.
The meaning of the relation (\ref{corresp}) is the following.  We choose as
an input any specific values of $m_t^{phy}$, $\tan \beta$, $m_{\tp}^{phy}$
and $m_{\bp}^{phy}$, and then we look for (``adjust'') $m_{\taup}^{phy}$
and $m^D_{\nutaup}$ ($E=1$ GeV) such that the q-l FD condition
(\ref{FDql2})
is satisfied,  
thus obtaining the l.h.s.~of (\ref{corresp}) as output. The see-saw
mechanism would then additionally provide us, similarly as in the
case of three generations (cf.~eq.~(\ref{see-saw1})), with an estimate
(upper bound)
for the physical neutrino mass on the r.h.s.~of eq.~(\ref{corresp})
\begin{equation}
m^{phy}_{\nutaup} ( \simeq \frac{(m^{D}_{\nutaup})^{2}}{M_{_{Majorana}}} )
\quad \stackrel{<}{\sim}  \quad
\frac{(m^{D}_{\nutaup})^{2}}{\Lb_{pole}}
=(m^{phy}_{\nutaup})^{u.b.} \ .
\label{see-saw2}
\end{equation}

These calculations, leading with the q-$\ell$ FD condition (\ref{FDql2})
from the l.h.s.~to the r.h.s.~of (\ref{corresp}) (and to (\ref{see-saw2})),
were here performed  with the 1-loop RGEs for the Yukawa couplings
for the case of four generations
(cf.~Appendix) in the 2HDSM(II). We considered the first and second
generations of fermions as essentially massless and ignored them, i.e.~their
mixing effects to the heavier fermions, in the RGEs. Furthermore, the
threshold for the evolution of the RGEs was taken to be $E_{thresh.} =
m_t^{phy}$, i.e.~the RGEs for the third and the fourth generation were
evolved only for energies $E \geq E_{thresh.} = m_t^{phy}$. For $E <
E_{thresh.}$, the physics was considered to be described by the effective
theory $SU(3)_c \times U(1)_{em}$ (without Higgs, W, Z, fourth generation
and the top)~\cite{ack}, within the corresponding evolution interval
[1 GeV,  $m_t^{phy}$]. We took in the RGEs the 2-loop solution for
the $SU(3)_c$ gauge coupling $\alpha_3$($E$), although it turned out that
the results do not depend appreciably on it\footnote{
We chose: $\alpha_3(M_Z) = 0.118$, corresponding to $\alpha_3$($E=34$ GeV) =
0.1387. The conclusions of this paper do not change if we take the
experimentally suggested upper bound
$\alpha_3$($E=34$ GeV) = 0.16  (see later).}.
For the light third generation masses we took\footnote{
For  simplicity,  we chose the third generation Dirac neutrino mass
$m^D_{\nu_{\tau}}$($E=1$ GeV) = 0.
It turned out that the results are virtually independent of the choice of
$m^D_{\nu_{\tau}}$($E=1$ GeV).}:  $m_b^{phy} \simeq 4.3$ GeV,
$m_{\tau}$($E=1$ GeV) $\simeq 1.78$ GeV.
To determine the physical masses of heavy
quarks, we used the (QCD-corrected) relation
\begin{equation}
m_q(E=m_q^{phy}) \simeq \frac{m_q^{phy}}
{1+ \frac{4}{3} \frac{\alpha_3(m_q^{phy})}{\pi}} \ .
\label{physmass}
\end{equation}

When calculating the correspondence (\ref{corresp}) with the
RGEs in this way (taking into account the q-$\ell$ FD
condition (\ref{FDql2})), we further impose four physical constraints:
\begin{equation}
m_{\tp}^{phy} \ , \ m_{\bp}^{phy} \ > \ m_t^{phy} \ ,
\label{c1}
\end{equation}
\begin{equation}
40 GeV \  \stackrel{<}{\sim} \ m^{phy}_{\nutaup} \  \stackrel{<}{\sim}
 \frac{(m^{D}_{\nutaup})^{2}}{\Lb_{pole}}  \ ,
\label{c2}
\end{equation}
\begin{equation}
\mbox{Yukawa couplings at electroweak scale are within perturbative range ,}
\label{c3}
\end{equation}
\begin{equation}
(\triangle \rho)_{heavy fermions~(h.f)} \ < \ 0.0076 \ .
\label{c4}
\end{equation}
The first two constraints are based on experimental data.
The third constraint was imposed to ensure that the  Standard Model
considered here (2HDSM(II)) is not manifestly
non-perturbative\footnote{
At $E \simeq  (0.7-0.75) \Lb_{pole}$, the 2-loop contribution to
the r.h.s.~of the RGEs (cf.~Appendix) for the Yukawa couplings
acquires approximately the same magnitude as the 1-loop contribution -
this estimate is based on the structure of the 2-loop RGEs for the MSM.}.
Specifically, we chose for (\ref{c3}) a rather generous constraint:
\begin{equation}
m_{\tp}^{phy} \ , \ m_{\bp}^{phy} \ \stackrel{<}{\sim} \ 0.5~ \Lb_{pole} \ .
\label{c3p}
\end{equation}
The 1-loop expression for the contributions to
$\triangle \rho$ from heavy fermions (in $\overline{MS}$
scheme) is~\cite{mv}
\begin{eqnarray}
(\triangle \rho)_{h.f.} & = & \frac{G_F \sqrt{2}}{16 \pi^2}
 \lbrace 3 K_{qcd} \lbrack m_t^2 +  m_{\tp}^2 + m_{\bp}^2
 - \frac{2 m_{\tp}^2 m_{\bp}^2}{(m_{\tp}^2-m_{\bp}^2)}
    \ln \frac{m_{\tp}^2}{m_{\bp}^2} \rbrack
 \nonumber\\
 & & + \lbrack m_{\taup}^2 + m_{\nutaup}^2
    - \frac{2 m_{\taup}^2 m_{\nutaup}^2}{(m_{\taup}^2-m_{\nutaup}^2)}
\ln \frac{m_{\taup}^2}{m_{\nutaup}^2} \rbrack \rbrace \ ,
\label{delrho}
\end{eqnarray}
where $m_t, \ m_{\tp}, \ m_{\bp}, \ m_{\taup}, \ m_{\nutaup}$ are
the physical masses, and $K_{qcd}$ is the QCD correction
parameter~\cite{Kqcd}
\begin{equation}
K_{qcd} \simeq  1 - \frac{(2 \pi^2+6)}{9 \pi} \alpha_3 \ .
\end{equation}
The fourth constraint (\ref{c4}) was obtained from an
essentially model-independent analysis of the LEP data~\cite{bv}.
Furthermore, the analysis of the experimental
evidence from $B - \bar B$ and $D - \bar D$ mixing, $\triangle m_K$,
$\epsilon_K$ and missing $E_T$ measurements at $p \bar p$ colliders
suggests bounds on $\tan \beta$~\cite{bhp}
\begin{equation}
0.5 \ \stackrel{<}{\sim} \ \tan \beta (\equiv \vev )
\ \stackrel{<}{\sim} \ 10 \ .
\label{interval}
\end{equation}
It is interesting that in the case of {\it three} generations we had obtained
$\tan \beta$ restricted between 0.53 and 2.1, once we imposed the q-l FD
at $\Lambda_{pole}$, the see-saw condition (\ref{see-saw1}) and
took the experimentally prejudiced values $(m_{\nu_{\tau}}^{phy})^{u.b.}
< 31$ MeV and $m_t^{phy}=(175 \pm 20)$ GeV~\cite{gccsk},\cite{moriond}.

The mass of the top quark has recently been measured by the CDF group at
Fermilab~\cite{CDF}: $m_t^{phy} \simeq (174  \pm 17)$ GeV. Therefore, we  first
took in our calculations $m_t^{phy}=160$ GeV, and varied $\tan \beta$.
When imposing  the first three constraints (\ref{c1})-(\ref{c3}) on the
obtained results,
we derived the allowed regions in the $m_{\bp}^{phy}$ vs.~$m_{\tp}^{phy}$
plane,
as depicted in Figs.~1, 2  and  3 for the cases $\tan \beta =$ 0.33, 1  and
3.5, respectively.
Some remarks are in order here. The lower (dotted) boundaries in the
figures originate from the heavy neutrino constraint (\ref{c2}), the
upper (full line) boundaries from the ``perturbative'' constraint
(\ref{c3}).
When further varying $\tan \beta$, we can obtain the possible region:
\begin{displaymath}
0.24 \ <  \  \tan \beta \ < \ 5.2 \ , \qquad (\mbox{for  }  m_t^{phy}=160
\mbox{ GeV})
\end{displaymath}
where the perturbative
constraint (\ref{c3p}) does not allow us to push $\tan \beta$ beyond
the above interval, since it then rules out the entire plane
$m_{\bp}^{phy}$ vs.~$m_{\tp}^{phy}$ (for $m_{\tp}^{phy}, m_{\bp}^{phy}
> m_t^{phy}$ = 160 GeV). Hence, the entire $\tan \beta$-interval
as permitted by the constraints (\ref{c1})-(\ref{c3}) is only within
$[0.24, ~ 5.2]$.

One surprising feature of Figs.~1, 2, 3 is
that the ``perturbative'' constraint (\ref{c3}) in both cases of
$\tan \beta =$ 0.33 (Fig. 1) and 3.5 (Fig. 3) turns out to be more restrictive
than in
the ``middle'' case of $\tan \beta = 1$ (Fig. 1). This is due to the fact that
the Yukawa coupling $g_{\tp}$ (, $g_{\bp}$) at low energies ($ E \simeq
m_t^{phy}$)  is  quite large\footnote{
$g_{\tp}$ (, $g_{\bp}$) at energies near $\Lb_{pole}$  is the dominant
quark Yukawa coupling in the cases of $\tan \beta < 1.2$ (, $> 1.3$,
respectively), for $m^{phy}_{\tp} \approx m^{phy}_{\bp}$}
in the case $\tan \beta = 0.33$ (, $3.5$, respectively), and
consequently it increases with energy rather quickly and $\Lb_{pole}$
is relatively small ($\Lb_{pole} \stackrel{<}{\sim} 10^4$ GeV).
In the case of $\tan \beta = 1$, on the other hand, both
$g_{\tp}$ and $g_{\bp}$ are relatively small at low energies and
consequently increase relatively slowly with energy, so that the
``perturbative'' constraint (\ref{c3}) is not as restrictive.

The most surprising thing is that, when we impose in addition the
``$\rho$-constraint'' (\ref{c4}) upon the regions in Figs.~1, 2, 3,
these regions become ruled out entirely. Namely, in the depicted
regions the calculations yield\footnote{
when taking for the QCD parameter $\alpha_3$($E=34$ GeV)  the experimentally
suggested upper bound 0.16, the values of $(\triangle \rho)_{h.f.}^{min}$
increase slightly, but less than one percent.}:
$(\triangle \rho)_{h.f.}^{min} =
0.0086$, $0.0155$, and $0.0084$, for $\tan \beta = 0.33$, $1$ and $3.5$,
respectively, all exceeding the maximum allowed value $0.0076$. This turns
out to be true for the entire interval $[0.24, ~5.2]$ for $\tan \beta$:
$(\triangle \rho)_{h.f.} \geq$ 0.0081.
The largest contribution to the values of $(\triangle \rho)_{h.f.}^{min}$
comes from $m_t^{phy}$ (0.0072), the rest predominantly from heavy leptons.
In the cases $\tan \beta \geq 3.5$, the minima for $(\triangle \rho)_{h.f.}$
are reached at the ``perturbative'' bound (where $m_{\tp}^{phy}
\simeq 0.5 \Lb_{pole}$).
All in all, the $\triangle \rho$-constraint
(\ref{c4}) cannot be satisfied in the regions allowed by the other
three constraints. Consequently, in the discussed case
of $m_t^{phy} = 160$ GeV, the q-$\ell$ FD condition (\ref{FDql2}) in the
flavor democracy--favored 2HDSM(II) model, together with the usual see-saw
mechanism,
effectively rules out the existence of the fourth generation.

When choosing higher $m_t^{phy}$ ($> 160$ GeV),
$(\triangle \rho)_{h.f.}^{min}$ becomes even bigger and the ``perturbatively
allowed'' interval for $\tan \beta$ becomes narrower.
Hence, we can argue that for $m_t^{phy} \geq 160$ GeV the q-$\ell$
FD requirement at
$E \simeq E_{trans.} \simeq \Lb_{pole}$ in the  flavor democracy--favored
2HDSM(II)
is {\it not} compatible with the existence of the fourth generation,
i.e.~this framework would practically rule out the fourth generation.
We also performed calculations at the experimental lower bound for
$m_t^{phy}$ (for $m_t^{phy} = 155$ GeV).
The ``perturbatively allowed'' interval for $\tan \beta$ is now
slightly larger: [0.23,~5.37]. However, the $(\triangle \rho)_{h.f.}^{min}$,
in the regions allowed by the first three conditions, is in this
case still above 0.0076, just reaching this value at $\tan \beta = 0.23, 5.37$,
as seen from Fig.~4.

If we used instead of the rather conservative ``perturbative'' constraint
(\ref{c3p}) a more restrictive one (i.e.~by replacing there $0.5 \Lb_{pole}$
by a smaller value), we would exclude the existence of the 4th generation
in the described framework even for the cases of $m_t^{phy}$ lighter
than 155 GeV.
Finally, we also investigated the effect of the CKM mixing by  introducing
$V_{\tp b}  \approx V_{\nutaup  \tau} \approx 0.2$ (at low energy), instead
of no 3-4 generation mixing,  and we found that the results
changed only for a fraction of one percent.   Therefore, the CKM mixing does
not
influence the results of this paper.

We conclude that for all $m_t^{phy} \geq 155$ GeV,
the existence of the fourth generation within the described
scenario (q-l FD at $\Lambda_{pole}$, and see-saw mechanism)
in the flavor democracy--favored 2HDSM(II) model
is practically ruled out. We also note  that
if we abandon the assumption of the see-saw mechanism, we cannot
rule out the existence of the fourth generation within the
discussed  flavor democracy framework. Namely, the low energy masses of $\taup$
and of the Dirac $\nutaup$ are always above 100 GeV and rather
close to each other, and we can choose
such $m_{\tp}^{phy} \approx m_{\bp}^{phy}$  that the contributions
of the fourth generation fermions to $(\triangle\rho)_{h.f.}$ are
practically zero, thus resulting in having no effects of the
$\triangle \rho$-constraint.

\vspace{0.4cm}

\noindent {\Large \bf Acknowledgement}

\vspace{0.4cm}

We would like to thank A.  Blondel and D.W. Kim for useful discussions on
$(\triangle\rho)_{h.f.}$.
The work of G.C.~was supported in part by Dortmund University and by
the Deutsche Forschungsgemeinschaft.
The work of C.S.K.~was supported
in part by the Korean Science and Engineering  Foundation,
in part by Non-Direct-Research-Fund, Korea Research Foundation 1993,
in part by the Center for Theoretical Physics, Seoul National University,
in part by Yonsei University Faculty Research Grant,  and
in part by the Basic Science Research Institute Program, Ministry of Education,
  1994,  Project No. BSRI-94-2425.

\newpage

\begin{appendix}

\noindent {\Large \bf Appendix: ``Type II Standard Model with two Higgs
doublets
and its RGEs''}

\vspace{0.4cm}

\setcounter{equation}{0}

In the 2HDSM(II), only one Higgs doublet ($H^{(u)}$) couples to the
``up-type'' right-handed fermions ($f_{uR}$) and is hence responsible
for their masses, and analogously the other Higgs
doublet ($H^{(d)}$) couples solely to the $f_{dR}$

\begin{eqnarray}
\L_{Yukawa} & = &          -  \sum_{i,j=1}^3 \lbrace
[( \bar q^{(i)}_L \tilde H^{(u)} ) q^{(j)}_{uR}U_{ij}^{(q)} + \mbox{h.c.}] +
[( \bar q^{(i)}_L        H^{(d)} ) q^{(j)}_{dR}D_{ij}^{(q)} + \mbox{h.c.}]
\rbrace
\nonumber\\
& &      -  \sum_{i,j=1}^3 \lbrace
[( \bar l^{(i)}_L \tilde H^{(u)} ) l^{(j)}_{uR}U_{ij}^{(\ell)} + \mbox{h.c.}] +
[( \bar l^{(i)}_L        H^{(d)} ) l^{(j)}_{dR}D_{ij}^{(\ell)} + \mbox{h.c.}]
\rbrace  \ ,
\end{eqnarray}
where $U^{(q)}$, $D^{(q)}$, $U^{(\ell)}$, $D^{(\ell)}$ are $4 \times 4$
Yukawa matrices in flavor basis (in the case of four generations), and we
use the notation

\begin{displaymath}
H^{(\alpha)} = {H^{(\alpha)+} \choose H^{(\alpha)o}} \ , \qquad
\tilde H^{(\alpha)} = i \tau_2 H^{(\alpha)\ast} \ , \
\langle H^{(\alpha)} \rangle_o = \frac{1}{\sqrt{2}}
{0 \choose {\it v}_{\alpha} } \qquad (\alpha = u, d) \ .
\end{displaymath}
\begin{displaymath}
q^{(i)} = {q^{(i)}_u \choose q^{(i)}_d} \ , \qquad
q^{(1)} = {u \choose d} \ , \ q^{(2)} = {c \choose s} \ , \
q^{(3)} = {t \choose b} \ , \ q^{(4)} = {\tp \choose \bp} \ ,
\end{displaymath}
\begin{equation}
l^{(i)} = {l^{(i)}_u \choose l^{(i)}_d} \ , \qquad
l^{(1)} = {\nu^D_e \choose e} \ , \ l^{(2)} = {\nu^D_{\mu} \choose \mu} \ , \
l^{(3)} = {\nu^D_{\tau} \choose \tau} \ , \
l^{(4)} = {\nu^D_{\taup} \choose \taup} \ .
\label{fermnotation}
\end{equation}

The corresponding 1-loop renormalization group equations (RGEs) for the
``squared'' Yukawa matrices
$Q^{(u)}$ ($ = U^{(q)}U^{(q)\dagger}$), $Q^{(d)}$
($ = D^{(q)}D^{(q)\dagger}$), $L^{(u)}$ ($= U^{(\ell)}U^{(\ell)\dagger}$)
and $L^{(d)}$ ($= D^{(\ell)}D^{(\ell)\dagger}$), in $\overline{MS}$
scheme are

\begin{eqnarray}
32 \pi^2 \frac{d}{dt}Q^{(u)} & = & 3 Q^{(u)^2} - \frac{3}{2}
  (Q^{(u)}Q^{(d)}+Q^{(d)}Q^{(u)}) + 2 Q^{(u)}(\Xi^{(q)}_u - A^{(q)}_u)
 \ , \nonumber\\
32 \pi^2 \frac{d}{dt}Q^{(d)} & = & 3 Q^{(d)^2} - \frac{3}{2}
  (Q^{(u)}Q^{(d)}+Q^{(d)}Q^{(u)}) + 2 Q^{(d)}(\Xi^{(q)}_d - A^{(q)}_d)
 \ , \nonumber\\
32 \pi^2 \frac{d}{dt}L^{(u)} & = & 3 L^{(u)^2} - \frac{3}{2}
  (L^{(u)}L^{(d)}+L^{(d)}L^{(u)}) + 2 L^{(u)}(\Xi^{(\ell)}_u - A^{(\ell)}_u)
 \ , \nonumber\\
32 \pi^2 \frac{d}{dt}L^{(d)} & = & 3 L^{(d)^2} - \frac{3}{2}
  (L^{(u)}L^{(d)}+L^{(d)}L^{(u)}) + 2 L^{(d)}(\Xi^{(\ell)}_d - A^{(\ell)}_d)
 \ ,
\label{RGE}
\end{eqnarray}
where
\begin{displaymath}
t = ln \left( \frac{2 E^2}{{\it v}^2} \right) \ ,
\end{displaymath}
\begin{displaymath}
\Xi^{(q)}_u = \Xi^{(\ell)}_u = Tr(3 Q^{(u)} + L^{(u)} ) \ ,
\end{displaymath}
\begin{displaymath}
\Xi^{(q)}_d = \Xi^{(\ell)}_d = Tr(3 Q^{(d)} + L^{(d)} ) \ ,
\end{displaymath}
\begin{displaymath}
A^{(q)}_u = \pi \left[ \frac{17}{3} \alpha_1+9 \alpha_2+32 \alpha_3 \right]
\ , \qquad A^{(q)}_d = A^{(q)}_u-4\pi\alpha_1 \ ,
\end{displaymath}
\begin{equation}
A^{(\ell)}_u = \pi [ 3\alpha_1+9\alpha_2] \ , \qquad
A^{(\ell)}_d = \pi [15\alpha_1+9\alpha_2] \ .
\end{equation}
Here, $E$ is the energy of probes,
$\alpha_1$, $\alpha_2$ and $\alpha_3$ are the usual Standard Model gauge
couplings corresponding to $U(1)_Y$, $SU(2)_L$ and $SU(3)_c$, respectively,
with $N_g = 4$ in the case of 4 generations.

\end{appendix}

\vspace{1cm}

\noindent{\large \bf Figure Captions:}

\vspace{0.6cm}

FIG.~1.~The region in the $m^{phy}_{\bp}$ vs.~$m^{phy}_{\tp}$ plane,
as allowed by the three constraints (\ref{c1})-(\ref{c3}). The lower
(dotted) boundary originates from the heavy neutrino constraint (\ref{c2}),
the other boundary (full line) from the ``perturbative'' constraint
(\ref{c3}) (i.e.  (\ref{c3p})). The figure is for the 2HDSM(II), with q-l FD
at $\Lb_{pole}$, $m^{phy}_t=160$ GeV, and $\tan \beta$($\equiv \vevskew$)=0.33.

\vspace{0.3cm}

FIG.~2.~As Fig.~1, but for $\tan \beta=$1.

\vspace{0.3cm}

FIG.~3.~As Fig.~1, but for $\tan \beta =$3.5

\vspace{0.3cm}

FIG.~4.~The $\triangle \rho$-contribution from heavy
fermions, i.e.~its minimal value ($(\triangle \rho)^{min}_{h.f.}$)
in the regions of the $m^{phy}_{\bp}$ vs.~$m^{phy}_{\tp}$ plane
that are allowed by the three constraints (\ref{c1})-(\ref{c3}),
as function of $\tan \beta$($\equiv \vevskew$). The full line is for
$m^{phy}_t=160$ GeV, the dotted for $m^{phy}_t=155$ GeV. Both
lines are above the maximal allowed value of 0.0076 (cf.~(\ref{c4})).
The figure is for the 2HDSM(II) with four generations, see-saw,
and q-l FD at $\Lb_{pole}$.

\end{document}